\documentclass[onecolumn,amsmath,amssymb,12pt,superscriptaddress,nofootinbib]{revtex5}
\pdfoutput=1
\usepackage{amsmath}
\usepackage{graphicx}
\usepackage{float}
\usepackage{color}

\begin{document}

\allowdisplaybreaks

\begin{titlepage}

\title{Quantum tunneling from paths in complex time}

\author{Sebastian F. Bramberger}
\email[]{sebastian.bramberger@aei.mpg.de}
\affiliation{Max Planck Institute for Gravitational Physics \\ (Albert Einstein Institute), 14476 Potsdam-Golm, Germany}
\author{George Lavrelashvili}
\email[]{george.lavrelashvili@tsu.ge}
\affiliation{Department of Theoretical Physics, A.Razmadze Mathematical Institute \\ I.Javakhishvili Tbilisi State University,
GE-0177 Tbilisi, Georgia}
\author{Jean-Luc Lehners}
\email[]{jlehners@aei.mpg.de}
\affiliation{Max Planck Institute for Gravitational Physics \\ (Albert Einstein Institute), 14476 Potsdam-Golm, Germany}

\begin{abstract}

\vspace{.3in}
\noindent
We study quantum mechanical tunneling using complex solutions of the classical field equations.  Simple visualization techniques allow us to unify and generalize previous treatments, and straightforwardly show the connection to the standard approach using Euclidean instanton solutions. We demonstrate that the negative modes of solutions along various contours in the complex time plane reveal which paths give the leading contribution to tunneling and which do not, and we provide a criterion for identifying the negative modes. Central to our approach is the solution of the background and perturbation equations not only along a single path, but over an extended region of the complex time plane. Our approach allows for a fully continuous and coherent treatment of classical evolution interspersed by quantum tunneling events, and is applicable in situations where singularities are present and also where Euclidean solutions might not exist. \\
\end{abstract}
\maketitle

\end{titlepage}

\tableofcontents

\vspace{2cm}

\section{Introduction}

Quantum tunneling follows as a direct consequence of wave mechanics, and can be calculated accurately by solving the Schr\"{o}dinger equation, if needed via numerical methods. The wavefunction on one side of a potential barrier is found to have (or to develop over time) non-zero support on the other side of the barrier too, resulting in a non-zero probability for observing a particle on the other side of the barrier even if its kinetic energy is lower than the potential energy of the barrier. Quantum tunneling thus provides a striking contrast with classical physics, in which the barrier is strictly unsurpassable given insufficient kinetic energy.

There exists a highly useful (semi-classical) approximation scheme to describe tunneling, which takes its roots in Feynman's path integral reformulation of quantum theory \cite{Feynman:1948ur}. This is the Euclidean instanton method, and its usefulness stems from the fact that it can be extended to quantum field theory \cite{Belavin:1975fg,Polyakov:1976fu,Coleman:1977py,Callan:1977pt} and even to gravity \cite{Coleman:1980aw}. We are particularly interested in the situation where the system under consideration (here a particle with position $x(t)$) can be described classically to a good approximation before and after the tunneling event. In Feynman's formulation, the transition amplitude from an initial state where $x(t_i)=x_i$ to a final state where $x(t_f)=x_f$ is given by the integral over all paths that link the two events,\footnote{We use natural units $c=1,$ $\hbar=1.$}
\begin{equation} \label{eqn:path-integral}
\langle x_f,t_f \mid x_i,t_i \rangle = {\mathcal N} \int_{x_i,t_i}^{x_f,t_f} D[x(t)] \, e^{iS}
\end{equation}
where ${\mathcal N}$ is a normalization factor and the action $S$ of a unit mass particle moving in a potential $V(x)$ is given by
\begin{equation} \label{eqn:action1}
S=\int \mathrm{d}t \left( \frac{1}{2}\dot{x}^2 - V(x) \right) \,,
\end{equation}
where we use the notation $\dot{}\equiv \frac{\mathrm{d}}{\mathrm{d}t}.$ Coleman has discussed in detail how, after performing a Wick rotation to Euclidean time, the associated path integral can be approximated by the saddle point method to determine the energy of a minimum of the potential and to deduce the decay rate out of a local minimum \cite{Coleman:1978ae}. The dominant field configuration contributing to the Euclidean-time path integral goes by the name of bounce or instanton, depending on the boundary conditions (bounces \cite{Coleman:1977py}
are used in the description of the decay of a metastable vacuum with $x(t_i)=x(t_f)=x_{min},$ while instantons, which correspond to ``half-bounces'',
describe either the splitting of energy levels for potentials with degenerate minima, or tunneling across a potential barrier \cite{Polyakov:1976fu}).
These solutions are finite action solutions of the Wick-rotated Euclidean equations of motion.

The standard description goes as follows: imagine a particle with insufficient kinetic energy to overcome a potential barrier. A good approximation is then to treat the particle classically as it runs up the potential barrier until it comes to a momentary stop on the slope of the potential when all its kinetic energy has been converted to potential energy. Here the possibilities bifurcate: the particle can either roll back down classically, or one can use the Euclidean time instanton solution to describe the tunneling of the particle to the other side of the barrier. The probability for this tunneling event to happen will be determined to leading order by the action of the instanton solution. The particle then emerges on the other side of the potential barrier with zero velocity, whence it can roll down the other side of the barrier classically. Thus the overall classical evolution in Lorentzian time is interrupted at an instant where the Euclidean time ``instanton'' solution is inserted.

As has often been discussed, this method works well but it is conceptually not very clear: how do we know that we can just put in the instanton solution between classical solutions? This procedure after all seems rather {\it ad hoc}. In the present paper we will attempt to answer that question by deriving a \emph{continuous} and generalized formulation of classical-to-quantum-to-classical transitions. Our approach will allow one to identify
the most relevant solutions for such transitions,
and will largely constitute both a justification and an extension of the instanton method. Conceptually our approach is clearer and more intuitive. Moreover, as we will argue, our methods will be useful in more complicated situations, in particular when gravity is included, when singularities are present and when Euclidean time instanton solutions might not exist.

\section{Tunneling via complex time paths}

\begin{figure}[h]
	\centering
	\includegraphics[width=0.9\linewidth]{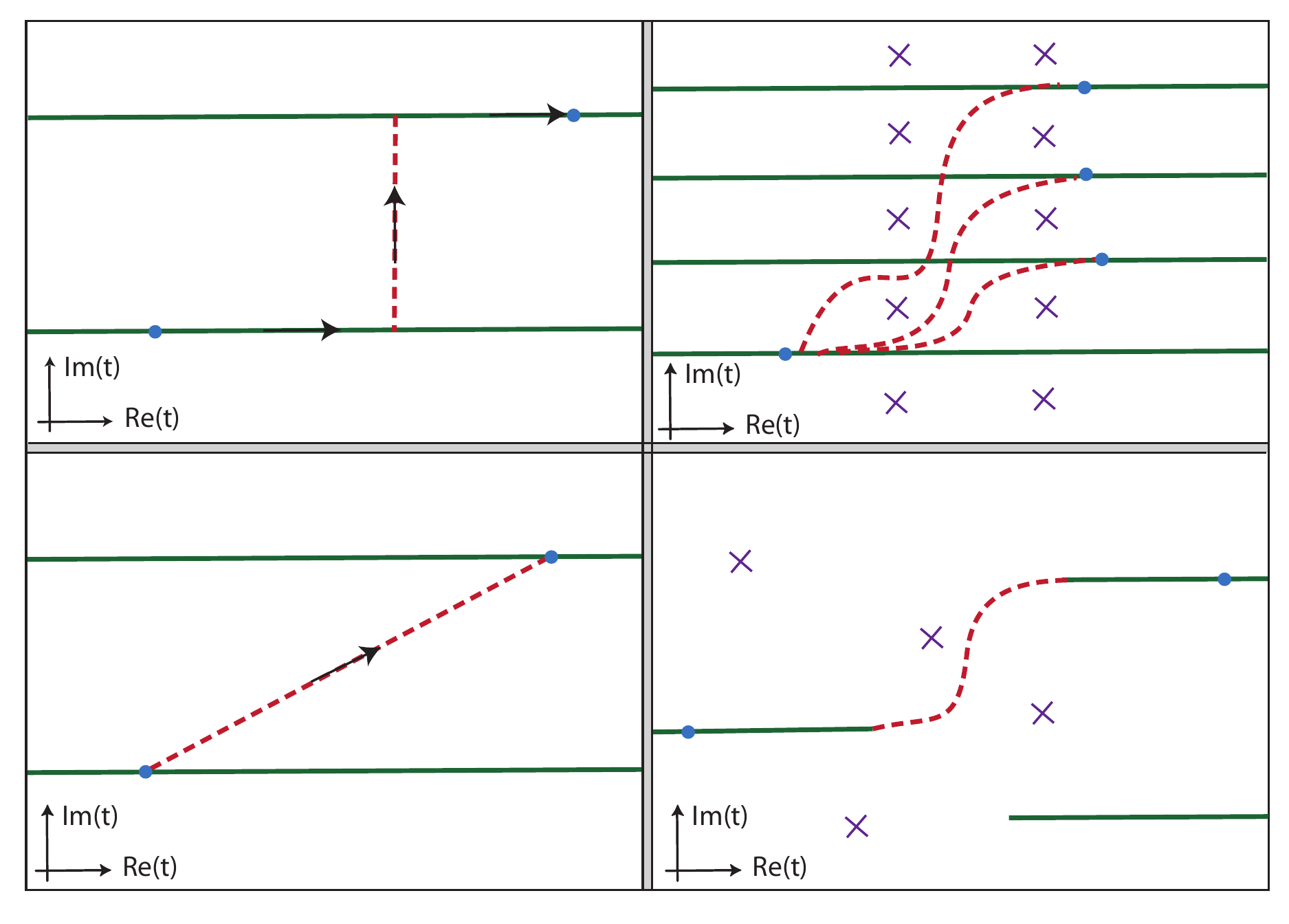}
	\caption{An overview of old and new approaches to describing classical to quantum and back to classical transitions. Green lines denote classical histories, while dashed red lines indicate Euclidean respectively fully complex tunneling paths. Blue dots show the location of initial and final conditions, while purple crosses indicate the location of singularities. For a full description of this figure see the main text.}
	\label{fig:Overview}
\end{figure}

Instead of employing only solutions of the equations of motion in either Lorentzian/real time or in Euclidean/imaginary time, we will consider solutions in terms of general complexified time. As discussed by many authors (in particular Levkov and Sibiryakov \cite{Levkov:2004ij},
Bender et al. \cite{Bender:2008fr,Bender:2009zza,Bender:2009jg,Bender:2010nu},
Dunne et al. \cite{Dumlu:2011cc,Behtash:2015zha,Behtash:2015loa}, Ilderton et al. \cite{Ilderton:2015qda}
and Turok \cite{Turok:2013dfa}) complex solutions of the classical field equations capture salient features of
quantum theory.
Moreover, as shown by Cherman and \"{U}nsal \cite{Cherman:2014sba} and by Turok \cite{Turok:2013dfa},
deformations of Euclidean time instanton solutions to a ``rotated'' time coordinate that approaches Lorentzian time seem to offer a sort of real time description of tunneling. In the present work we unify and extend these approaches. We arrive at the following picture -- see Fig. \ref{fig:Overview}.

Purely classical evolution corresponds to evolution along a line parallel to the real time axis, with all field values (and derivatives) taking real values. These are the green lines in Fig. \ref{fig:Overview}. It is important to realize that, if the fields take real values, it is not necessary for the evolution to be represented exactly on the real time axis, rather any line parallel to it will do equally fine since the differential $\mathrm{d}t$ is also real on that line, and hence the momenta will also be real. The {\it upper left} panel then illustrates the picture suggested by standard instanton methods: from a classical solution one can tunnel via a Euclidean time instanton solution (indicated by a dashed red line) to a different classical history. The idea here is that in the classically forbidden region, the leading approximation to the transition amplitude \eqref{eqn:path-integral} is given by a saddle point of the Wick-rotated ($\tau = - i t$) Euclidean action
\begin{equation}
S_E = - i S = \int \mathrm{d}\tau \left(\frac{1}{2}(x_{,\tau})^2 + V(x) \right)\,,
\end{equation}
that is to say by a classical solution of the Euclidean equations of motion with finite action $S_{E,instanton}$. Moreover, to leading approximation the probability for this tunneling to take place is given by the factor $e^{-2 S_{E,instanton}}.$ We can picture this sequence of events in the complexified time plane, as shown in the figure: two classical histories in real time are joined by a Euclidean solution mediating the tunneling. The fact that the transition is classically forbidden is reflected in a shift along the Euclidean time axis. As soon as one has this picture in mind, it becomes clear that the path taken in the complex time plane may also be deformed, as long as it does not pass any singularities of the solutions of the {\it complexified} classical equations of motion. This brings us to the {\it lower left} panel, which illustrates the approach of Cherman-\"{U}nsal \cite{Cherman:2014sba} where the tunneling path is rotated so as to be aligned more and more with the classical histories it is meant to join. In \cite{Cherman:2014sba} only the tunneling part is considered, and rotations arbitrarily close to the real line are advocated, but as our graph indicates, the boundary conditions will limit to what extent such a rotation is feasible. The {\it upper right} panel illustrates the point of view advocated here: a classical history can tunnel to various other classical histories via various paths in the complex time plane. These paths are equivalent as long as their deformations do not encircle singularities (marked by purple crosses) and of course as long as they respect the specified initial and final conditions (in the figure we show three paths with different final conditions). What is not shown here is that paths that differ in how they circle singularities can take the evolution onto a different sheet of the solution function, and on this new sheet both the singularities and the loci of classical histories may differ from other sheets. We will discuss this in more detail in section \ref{Sec:Examples} and present an example illustrating these concepts. The {\it lower right} panel shows a situation in which our method will be of clear advantage over existing ones: there exist classical histories which cannot be joined via purely Euclidean time instanton solutions. Moreover a variety of singularities are present. In this case our method will nevertheless allow one to determine which complex time paths can mediate a quantum transition between different classical histories.

The crucial question we have not discussed yet is which paths actually provide the leading contribution to tunneling and which do not. The standard instanton method employs a single path, but how do we know that this is the dominant/relevant path? Evidently, by Cauchy's theorem we can deform a path in the complex time plane as long as it does not cross any singularities. Such deformed paths are entirely equivalent to the original one, and should not be counted multiple times. However, in general singularities will be present, and then there exist inequivalent paths that encircle the singularities in various ways. Should we then sum over all possible (inequivalent) complex paths between fixed initial and final conditions? As we will now argue, the answer to this question is ``no''. Not all such paths are equally relevant for tunneling, and we will now identify a criterion for identifying the relevant path(s).

The crucial notion here is to look at {\it fluctuations} around all possible interpolating paths. For purely Euclidean instantons this analysis was first performed by Callan and Coleman in \cite{Callan:1977pt}. Consider again the saddle point approximation. Around the saddle point, where the solution to the Euclidean equation of motion is denoted by $x_{cc},$ to quadratic order the action can be approximated by
\begin{eqnarray}
S_E[x,x_{,\tau}] = S_E[x_{cc}]  + \frac{1}{2} \int_{x_i,\delta x(\tau_i)=0}^{x_f,\delta x(\tau_f)=0} \mathrm{d}\tau \left(  (\delta x_{,\tau})^2 + V''(x_{cc}) (\delta x)^2 \right) + \cdots,
\end{eqnarray}
where $V''=V_{,xx},$ and where the term linear in $\delta x$ vanishes precisely because we expand around an extremum. Given the boundary conditions on $\delta x$ (vanishing at the endpoints), we can expand any fluctuation into a complete set of eigenfunctions of the fluctuation operator,
\begin{equation}
\delta x = \sum_n c_n \delta x_n \,,
\end{equation}
where $\int \mathrm{d}\tau \, \delta x_n \, \delta x_m = \delta_{nm},$ and obeying the eigenvalue equation
\begin{equation}
\left[ - \frac{\mathrm{d}^2}{\mathrm{d} \tau^2}+ V''(x_{cc})\right] \delta x_n = \omega_n \delta x_n\,,
\label{eqn:pert-eqn}
\end{equation}
where the $\omega_n$ are the (real) eigenvalues. The integral above then turns into simple Gaussian integrals, which can be performed to yield the approximation\footnote{When zero modes are present, they must be treated separately. A proper inclusion of the zero modes results in an additional prefactor which is irrelevant for our discussion \cite{Callan:1977pt}.}
\begin{equation}
\langle x_f,t_f \mid x_i,t_i \rangle = {\mathcal N} \int_{x_i,t_i}^{x_f,t_f} D[x(t)] \, e^{iS} \sim e^{-S_E(x_{cc})} \frac{1}{\sqrt{\Pi_n \omega_n}}\,.
\end{equation}
The Gaussian integrals result in a prefactor that involves the square root of the product of eigenvalues of the fluctuation operator. If all eigenvalues are positive, then any fluctuation around the saddle point solution will increase the action, resulting in a lower probability. In this case we have found the dominant path. On the other hand, if some of the eigenvalues are negative, then there exist fluctuations that can lower the Euclidean action. Such solutions are thus not actual extrema and must be discarded\footnote{A Euclidean solution which describes the {\it decay} of a metastable vacuum, i.e. a bounce, has exactly one negative mode, which justifies the decay picture. Here we are interested in tunneling solutions, i.e. instantons, which should have at most zero modes in their spectrum of linear perturbations. As resurgence theory showed, the  contributions of higher order (complex) saddles can
sometimes play an important role e.g. in order to obtain consistency
of the semiclassical expansion with the supersymmetry algebra \cite{Behtash:2015loa}.
Here we will be identifying solutions providing the {\it leading} contribution
to tunneling, i.e. having the least Euclidean action and no negative modes.
We call such solutions "relevant", as opposed to "irrelevant" higher order saddles, which we discard in the leading approximation.}. (For a related discussion see \cite{Coleman:1987rm}.) How do we know whether negative modes exist? After all, it might be difficult to find the associated eigenfunctions numerically. Here the nodal theorem helps (see e.g. \cite{aq95} and references therein): we can solve the perturbation equation \eqref{eqn:pert-eqn} for the zero eigenvalue $\omega=0,$ with the boundary conditions $\delta x(\tau_i)=0,\, \delta x_{,\tau}(\tau_i) = \pm 1$ (since this is not necessarily an eigenfunction we do not care about normalizability and can in principle choose $\delta x_{,\tau}$ to take any non-zero value). The number of nodes of the corresponding solution, which we refer to as the {\it perturbation function}, will tell us the number of negative modes\footnote{Think about the energy eigenfunctions in a potential well: with each increasing eigenvalue an additional node is present. Hence if the solution with zero eigenvalue has $n$ nodes, there must exist $n$ eigenfunctions with lower, i.e. negative, eigenvalues.}. In this way we can determine whether we have found
the most relevant tunneling solution
without having to find the eigenfunctions and eigenvalues of \eqref{eqn:pert-eqn} explicitly.

Now we want to adapt this argument to the case where the paths under consideration are complex. In fact, we will retain the Euclidean formulation, but where one should now consider both the Euclidean time coordinate and the fields to be complexified (it may appear baroque to rotate to Euclidean time before complexifying, but this avoids the use of slightly awkward factors of $i$ -- we will discuss how to get back to Lorentzian time below). By analytically continuing, the eigenfunctions will become complex but the eigenvalues $\omega_n$ remain real as these are simply constants. The problem is that the nodes in the analytically continued perturbation functions will in general disappear, and thus it looks like we might lose our simple criterion for determining which paths are relevant and which are not. However, we can find a resolution of this issue by thinking about the nodes in a little more detail: if a node is present in the Euclidean zero-eigenvalue perturbation function $\delta x_0$ at $\tau_0$ then because of the boundary conditions we can expand the perturbation function between $\tau_i$ and $\tau_0$ using purely $\sin$ functions,
\begin{equation}
\delta x_0 = \sum_k c_k \sin \left(\frac{k\pi}{(\tau_0-\tau_i)}(\tau-\tau_i)\right)\,,
\end{equation}
where $k \in \mathbb{N}$ runs over integer values. Now imagine that we deform the solution path by shifting it along the Lorentzian time direction by a constant amount $\Delta \tau=i \Delta t,$ where $\Delta t \in \mathbb{R}.$ Then along the Lorentzian time direction starting from the node at $\tau_0$ we have
\begin{equation}
\sin \left(k\pi + i \Delta t \right) = \sin\left(k\pi\right)\cos\left(i\Delta t\right) + \cos\left(k\pi \right)\sin\left(i \Delta t\right) = \mp \sin \left(i \Delta t\right) = \mp i \sinh\left( \Delta t \right)\,. \label{node-location}
\end{equation}
From a node, and along the Lorentzian time direction, the change in the perturbation function will therefore be purely imaginary! This implies that if we look at the real part of the zero-eigenvalue perturbation function it will still contain a node. Thus we can essentially retain the same criterion for deciding whether solutions are relevant or not as in the pure Euclidean case, with the proviso that we must look only at the real part of the perturbation function. There is one possible caveat: could the complex perturbation function accidentally develop a zero in its real part, i.e. a zero not related to an actual node? This certainly seems conceivable, but in practice it is easy to avoid any ambiguity: the above arguments imply that if one solves for the zero-eigenvalue perturbation function over an extended region in the complex time plane, then there will be an entire line of zeros associated with an actual node. Such a line of zeros is thus the unmistakeable signature of solutions that must be discarded. Furthermore, the freedom to deform the contour in the complex time plane implies that one can always deform the solution path such that it crosses such a line of zeros and then comes back. Our criterion for finding the dominant tunneling solution may therefore be stated more carefully as follows: \emph{if the real part of the zero eigenvalue perturbation function unavoidably crosses a line of zeros, this signals the presence of a negative mode and the solution must be discarded. If the real part of the zero eigenvalue perturbation function does not cross any such line of zeros, the solution is relevant to tunneling.}

We may now go back to Lorentzian time and reformulate this calculation in terms of complexified real time $t.$ The zero-eigenvalue perturbation function $\psi$ must then satisfy the following equation of motion and boundary conditions
\begin{equation}
\left[ \frac{\mathrm{d}^2}{\mathrm{d} t^2}+ V''(x_{cc})\right] \psi = 0, \qquad \psi(t_i)=0, \quad \dot\psi(t_i)=\pm i\,.
\label{eqn:pert-eqn-Lorentzian}
\end{equation}
Since we have also transformed the boundary conditions in accordance with the change of time coordinate, our criterion above remains unchanged and we must look for lines of nodes of $Re(\psi),$ as we will do in section \ref{Sec:Examples}. The presence of such a line of zeros will imply that a particular solution must be discarded, while a solution without any such nodes in its perturbation function will be relevant to tunneling.

A couple of additional comments: the existence of  solutions with more and more negative modes is reminiscent of gravitational oscillating bounces \cite{Hackworth:2004xb,Lavrelashvili:2006cv,Lee:2011ms,Battarra:2012vu,Lee:2014ula}, which also seem to represent excited states
not contributing dominantly to the description of vacuum decay.
The existence of such oscillating instantons is usually explained by arguments about Hubble friction and anti-friction, while here we will see that qualitatively similar solutions can exist even in the absence of gravity. Further works discussing the importance of negative modes in quantum tunneling include \cite{Lavrelashvili:1985vn,Tanaka:1992zw,Lavrelashvili:1999sr,Khvedelidze:2000cp,Gratton:2000fj,Dunne:2006bt,Battarra:2013rba,Lee:2014uza,Koehn:2015hga}.

\begin{figure}[]
	\centering
	\includegraphics[width=0.5\linewidth]{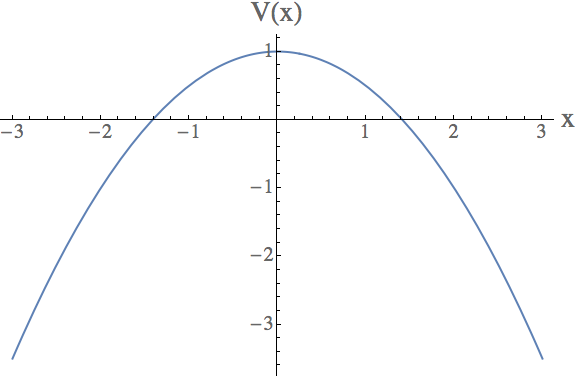}
	\caption{Plot of an inverted harmonic oscillator potential, $V(x)=1-\frac{1}{2}x^2$.}
	\label{fig:inv-sho-pot}
\end{figure}

\section{Examples} \label{Sec:Examples}

The discussion so far might have seemed rather generic and abstract.
We will now illustrate the ideas discussed above with concrete examples.

The core of our approach is the solution of the background and perturbation equations over an extended region
of the complex time plane, and the visualization of the results by means of relief plots.
Let us briefly describe how exactly this is done.
First we note that the Lorentzian action \eqref{eqn:action1} can be written in a reparametrization invariant way
\begin{equation}
S=\int n \, \mathrm{d}\lambda \left( \frac{1}{2 n^2}(x_{,\lambda})^2 - V(x) \right) \,,
\end{equation}
where $n(\lambda)$ is the (complex) ``lapse function'' and $\lambda$ is a (real) parameter. Choosing a particular form for $n(\lambda)$ then allows one to follow a specified path in the complex time plane. For instance, $n(\lambda)=1$ corresponds to evolving along the Lorentzian time direction, while $n(\lambda)=i$ corresponds to the Euclidean direction -- more general choices of $n(\lambda)$ will allow evolution along any desired curve in the complex time plane.

We solve the equations of motion, starting from purely classical boundary conditions. The solution along the Lorentzian time axis then gives the classical solution, with real (conserved) energy, which in the case of a barrier potential means the solution that rolls up the potential until all kinetic energy is converted to potential energy; subsequently the particle simply rolls back down the potential. From this reference solution we branch out in both perpendicular directions, integrating the equation of motion as we go along while periodically sampling the values thus obtained. By repeating this procedure, and shifting the integration path by a small amount every time, we obtain the solution over a dense grid of points in the complex time plane. If no singularities are present, this prescription already gives us the full solution over the required time domain (with a resolution limited by the accuracy of the numerical computation). If singularities are present, then the reference path can be deformed repeatedly so as to encircle the singularities in various ways (where now branch cuts automatically appear ``behind'' the singularities after branching out from the reference path), until all possible paths are explored. A detailed example of this latter situation will be presented in section \ref{Sec:Sing}. The same procedure can then also be repeated for the perturbed equation of motion, imposing the boundary conditions specified in \eqref{eqn:pert-eqn-Lorentzian}. We then employ relief plots to visually represent the real and imaginary parts of the background and perturbation solutions. The three examples below will illustrate this procedure.

\subsection{Inverted harmonic oscillator} \label{Sec:IHO}

As a first example consider a particle moving in an inverted harmonic oscillator potential (see Fig. \ref{fig:inv-sho-pot})
\begin{align}
V(x) = -\frac{1}{2}\Omega^2 x^2 + V_0\,,
\end{align}
where $\Omega, V_0$ are constants. This potential is unbounded from below, but one might imagine that it gets deformed so as to develop a minimum at large field values -- in any case, we are just interested in energy differences here. The potential has the advantage that analytic solutions to both the background equation of motion ($\ddot{x}=\Omega^2 x$) and perturbation equation ($\ddot{\psi} =\Omega^2 \psi$) can be found easily. They are both given by
\begin{equation}
x(t),\psi(t)=c_1 e^{\Omega t} + c_2 e^{-\Omega t}, \qquad t \in \mathbb{C},
\end{equation}
where $c_1, c_2$ are complex integration constants to be determined. The solution is exponential when $t$ is purely real and periodic when $t$ is imaginary. If we choose the origin of time $t=0$ to correspond to the moment just before tunneling, then the background solution is
\begin{equation}
	x(t)=c \cosh{\Omega t}, \qquad t \in \mathbb{C}
\end{equation}
where the constant $c$ is the particle's location at the classical turnaround at $t=0.$

\begin{figure}[h]
	\centering
	\includegraphics[width=\linewidth]{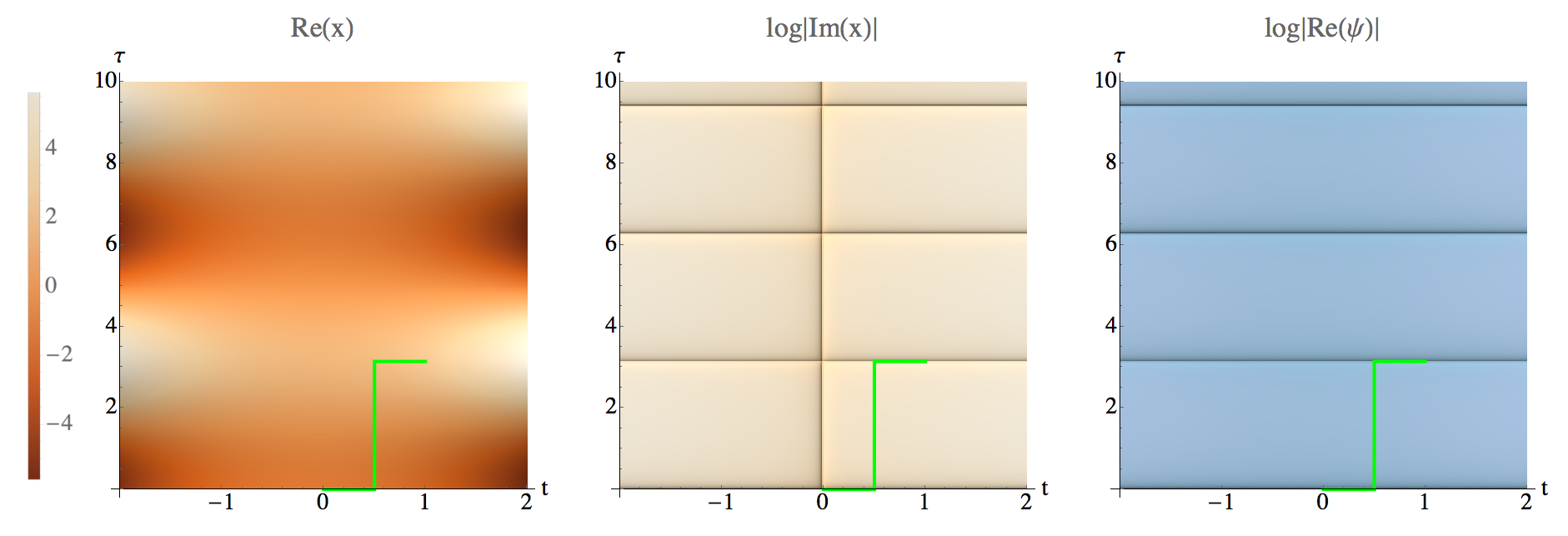}
	\caption{Relief plots of the background (left and center panels) and perturbation (right panel) solutions in the inverted harmonic oscillator potential. The horizontal axis indicates the real time direction, while the vertical axis indicates the imaginary/Euclidean time direction. Darker colors represents smaller (more negative) values while brighter colors represents larger (more positive) values. Therefore in the center and right images the black lines show the regions where $\mathrm{Im(x)}$ and $\mathrm{Re(\psi)}$ are zero, indicating the regions of classicality along with the Euclidean instanton solutions (center panel), and the locations of nodes (right panel) respectively. This means for example that along a path following a black line in the center panel, the field value will be purely real. Similarly, along a path following a black line in the right panel, the perturbation function will be purely imaginary. The usefulness of these graphs stems from the fact that they allow one to see the behavior of not just a single solution, but of all solutions (on a particular sheet) in an extended region of the complex time plane. The green line indicates a particular path which is further inspected in figure \ref{fig:inv-sho-pot-EoM-path1}.}
	\label{fig:inv-sho-pot-EoM-Grid}
\end{figure}

\begin{figure}[h]
	\centering
	\includegraphics[width=\linewidth]{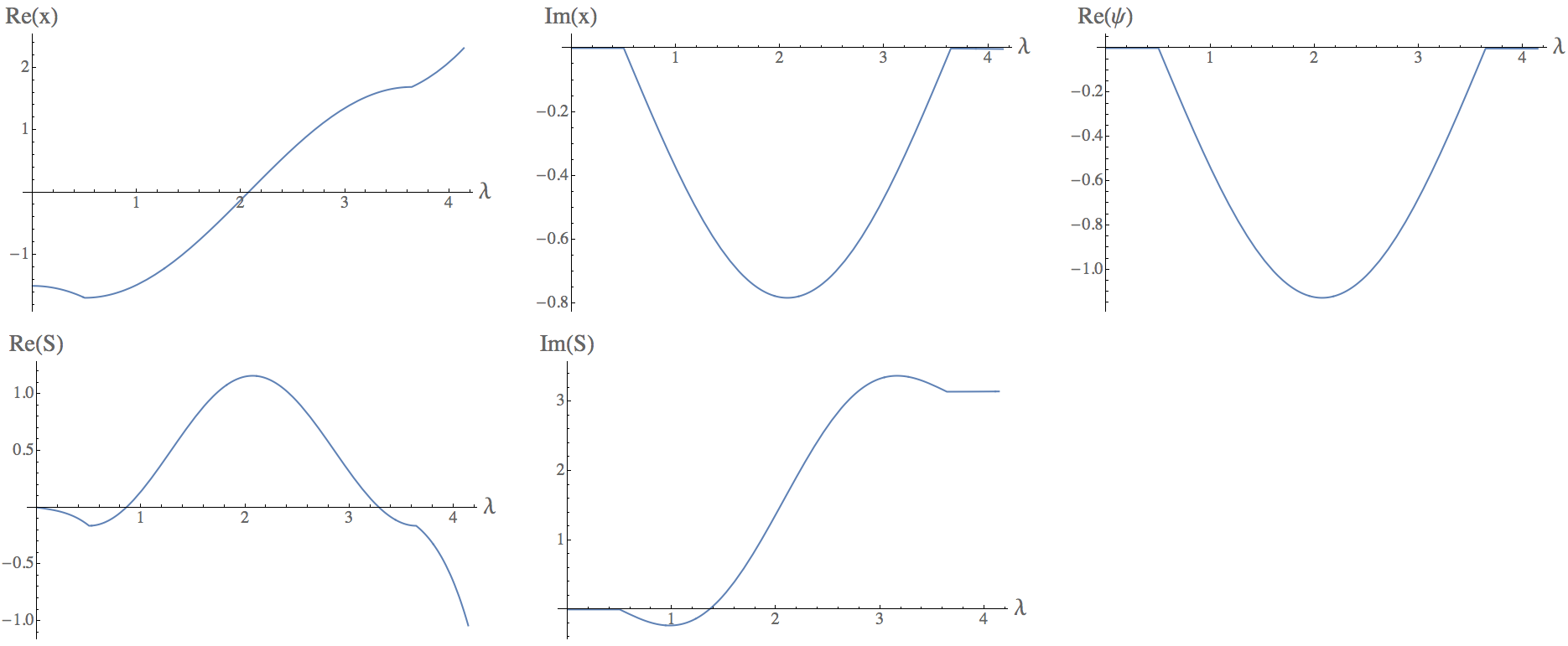}
	\caption{Field values and action for the tunneling path drawn by the green line in Fig. \ref{fig:inv-sho-pot-EoM-Grid}. Note that this is an actual tunneling path, with $Re(x)$ interpolating between two different sides of the potential and $Im(x)$ returning to zero after tunneling. The imaginary part of the action reaches a constant after tunneling, and this value will (to leading order) determine the probability for this tunneling event to take place. As required, the real part of the perturbation function $\psi$ does not present any nodes.}
	\label{fig:inv-sho-pot-EoM-path1}
\end{figure}

Even though for this particular case analytic solutions are available, this will of course in general not be the case. For this reason we will in general solve the equations of motion numerically over an extended region of the complex time plane. Here we do this in Fig. \ref{fig:inv-sho-pot-EoM-Grid} (where we have taken $c=-3/2$), so that we can directly compare our numerical methods with the analytic results. As explained above, the figures are obtained by solving the equations of motion over a dense grid of points in the complex time plane, and then representing the results with relief plots. Then, by taking a look at $\log|\mathrm{Im(x)}|$ one can immediately identify where the solutions are purely real: note that $\log|\mathrm{Im(x)}|$ blows up to large negative values for small imaginary values of $x$ and thus the locations where $x$ has a tiny or zero imaginary part will be represented by very dark colors. In this way the regions of classicality become obvious by inspection. There are infinitely many lines parallel to the real time axis along which the solution is real and classical. Tunneling then corresponds to considering complex time paths that join different such (horizontal) classical solutions by traversing regions of non-classicality. One path is singled out in the graph, namely the Euclidean instanton solution, which is the vertical dark line
in Fig. \ref{fig:inv-sho-pot-EoM-Grid}, center panel.
This solution stands out since the field values are purely real along it. However, in our approach this path is now not any more fundamental than other paths through the complex time plane. An example of a possible tunneling path is drawn in Fig. \ref{fig:inv-sho-pot-EoM-Grid}, with the evolution of the field and action along this path shown in Fig. \ref{fig:inv-sho-pot-EoM-path1}.\footnote{Note that we only need to consider tunneling paths in one direction along the Euclidean time direction, namely the direction corresponding to the correct Wick rotation. In practice this direction can be identified by the fact that quantum tunneling is suppressed compared to classical evolution, i.e. that the imaginary part of the action is positive.} After tunneling, the solution is given by
\begin{equation}
x(t)= c \cosh \left(i \pi+ \Omega t \right) = - c \cosh \Omega t, \qquad t \in \mathbb{C}
\end{equation}
and it is classical again, as it should.

We have also plotted (the real part of) the perturbation function, which satisfies \eqref{eqn:pert-eqn-Lorentzian} and in the present case is given by
\begin{equation}
\psi(t)=\frac{i}{\Omega}\sinh{\Omega t}, \qquad t \in \mathbb{C}\,.
\end{equation}
The right panel in Fig. \ref{fig:inv-sho-pot-EoM-Grid} shows the zeros of the real part of the perturbation function. As expected, these nodes are distributed along continuous lines. Paths that join two adjacent lines of classicality can avoid crossing any node, and hence these paths are the relevant ones for tunneling. By contrast, a tunneling path joining two lines of classicality that are separated by additional lines of classicality in between will contain nodes, and hence must be discarded. This simple example thus illustrates the main concepts advocated in the previous section.

\subsection{Inverted Higgs potential}

\begin{figure}[h]
	\centering
	\includegraphics[width=0.5\linewidth]{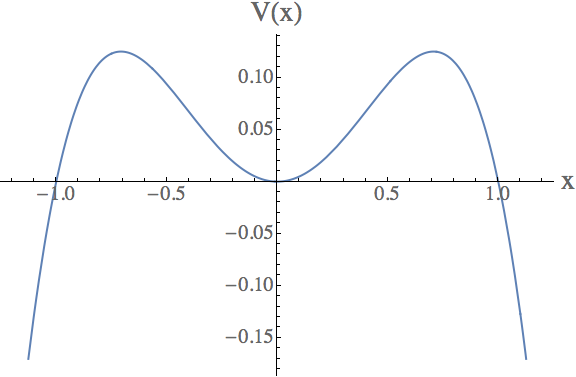}
	\caption{Plot of the inverted Higgs potential $V(x) = \frac{1}{2} x^2 - \frac{1}{2} x^4$.}
	\label{fig:inv-higgs-pot}
\end{figure}

\begin{figure}[h]
	\centering
	\includegraphics[width=\linewidth]{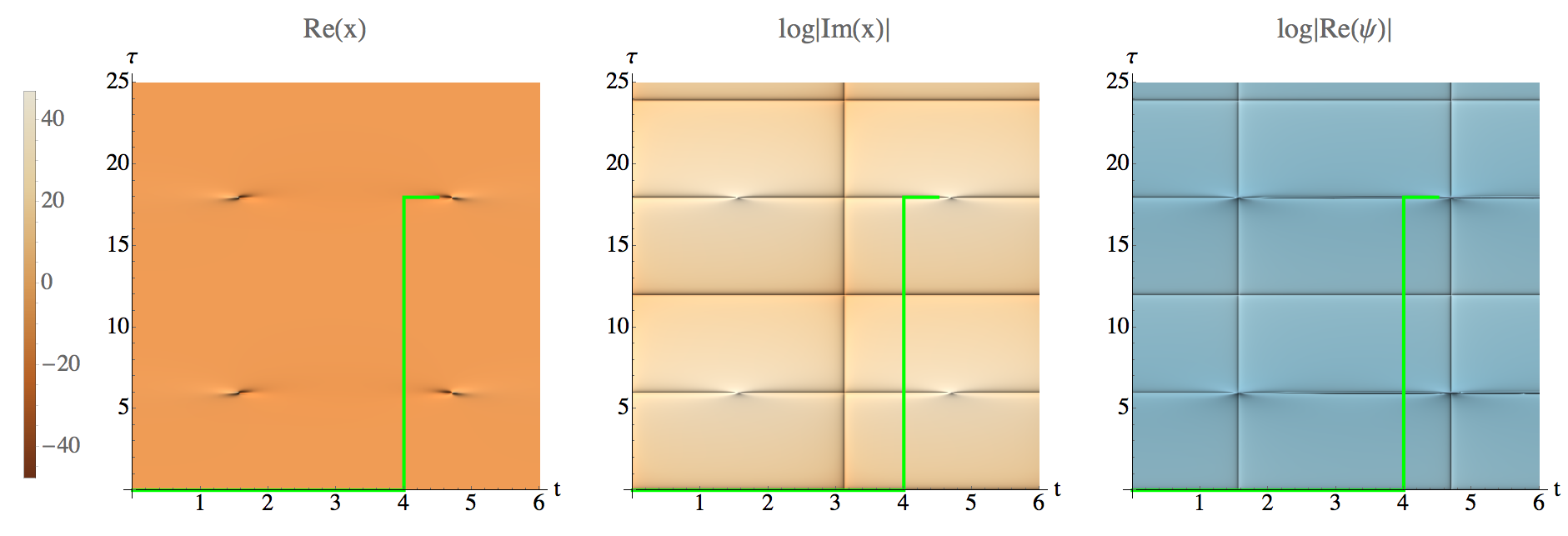}
	\caption{A figure analogous to Fig. \ref{fig:inv-sho-pot-EoM-Grid}, but for the inverted Higgs potential. The plots are obtained with the initial condition that the particle is released at $x=10^{-2}$ with zero velocity.}
	\label{fig:inv-higgs-pot-path2-grid}
\end{figure}

\begin{figure}[h]
	\centering
	\includegraphics[width=\linewidth]{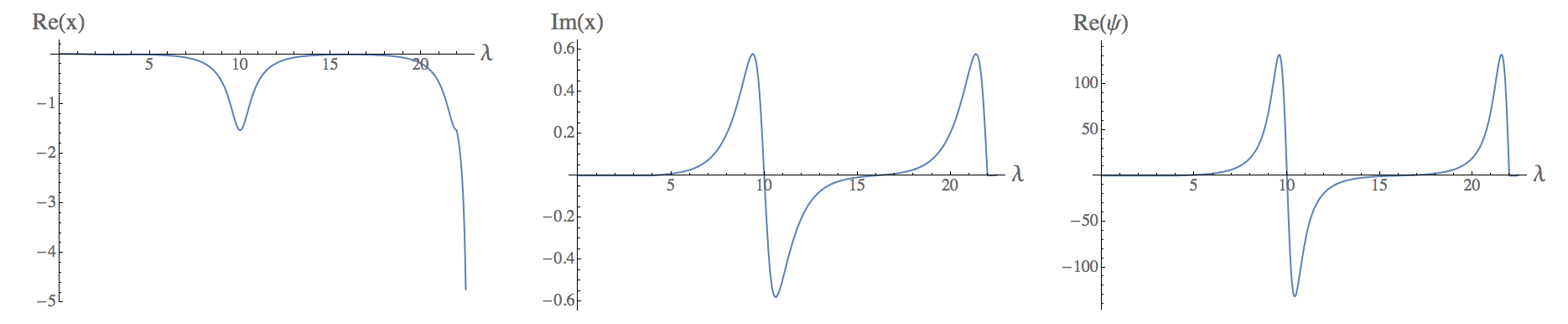}
	\caption{An example of an irrelevant solution. The path chosen here is indicated by the green line in Fig. \ref{fig:inv-higgs-pot-path2-grid}. The perturbation function contains two nodes, indicating that there exist perturbations of this solution that increase the probability.}
	\label{fig:inv-higgs-pot-path2}
\end{figure}

Next, consider a particle moving in an inverted Higgs potential,
\begin{align}
V(x) = \frac{1}{2} x^2 - \frac{1}{2} x^4\,, \label{IHPot}
\end{align}
which has also been studied by Turok \cite{Turok:2013dfa}. The potential is shown in Fig. \ref{fig:inv-higgs-pot}. The general solution is \cite{Turok:2013dfa}
\begin{equation}
x(t)=-\frac{1}{\sqrt{1+m}}\frac{1}{sn(t/\sqrt{1+m}|m)}, \qquad t \in \mathbb{C}\,,
\end{equation}
where $sn$ denotes the doubly periodic Jacobi $sn$ function, and the order $m$ of the function determines the energy of the solution, $2E=m/(1+m)^2$. A particularly simple limit is obtained by setting the energy to zero, $m=0,$ in which case the solution is
\begin{equation}
x(t)=-\frac{1}{\sin t}, \qquad t \in \mathbb{C}\,,
\end{equation}
where for this solution the particle is at negative infinity at $t=0$ and reaches the turn-around/tunneling location $x=-1$ at $t=\frac{\pi}{2}.$ The perturbation function, satisfying the required boundary conditions at $t=\frac{\pi}{2},$ is given by
\begin{equation}
\psi(t) = i \, \frac{\cos(t)}{\sin^2(t)}, \qquad t \in \mathbb{C}\,.
\end{equation}
Plots of the background solution and the real part of the perturbation function are shown in Fig. \ref{fig:inv-higgs-pot-path2-grid}, for the case of a small positive energy. In all plots the double periodicity is immediately apparent. The dark spots in the left panel show the regions where the particle rolls to large field values. The center panel indicates the lines of classicality. Once again we have an infinite number of such lines parallel to the real time axis (there is also a vertical line in the middle along which the field is real -- this is the Euclidean instanton solution). Possible tunneling paths then join two such horizontal lines. As the right panel shows, joining two adjacent lines will not result in having nodes in the perturbation function, and such paths thus contribute to tunneling. Lines of classicality that are further separated in the Euclidean time direction are also separated by lines of nodes, and hence the corresponding tunneling solutions must be discarded. For illustration, an example of such an irrelevant solution is given in Fig. \ref{fig:inv-higgs-pot-path2}.

For completeness we should discuss the vertical line of nodes in the right panel of Fig. \ref{fig:inv-higgs-pot-path2-grid}. This line is located at the position in real time where the particle reaches the potential minimum at $x=0$ (and it is a direct consequence of the $\cos$ expansion of the background solution that one can perform around that point). It thus divides the evolution into regions left or right from the local minimum of the potential, and in this manner divides it into regions with the possibility to tunnel across either the left or right barrier. If we imagine having a particle on the left of the local minimum (i.e. at $x<0$) but say we want to evaluate the transition amplitude to emerge on the far side of the right potential barrier, then we may follow the classical evolution from $x<0$ to $x>0$ first, and then tunnel across the right potential barrier. In this sense this vertical line of nodes is avoidable and therefore does not obstruct a contribution to the path integral.
	
\subsection{Potential barrier with singularities} \label{Sec:Sing}
		
Our final example is also the most interesting one, namely a potential hill of the form		
\begin{align}
		V = \frac{1}{x^2 + 1}\,.
\end{align}
		
\begin{figure}[h]
			\centering
			\includegraphics[width=0.5\linewidth]{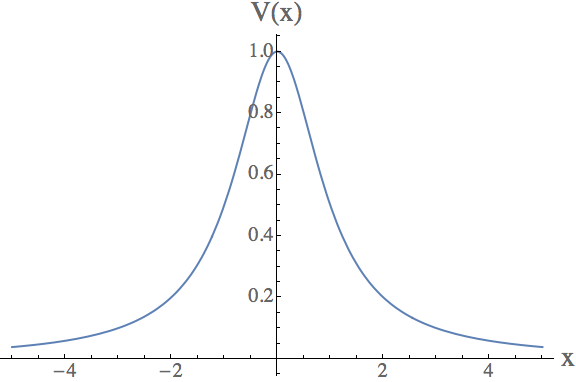}
			\caption{A potential hill $V = \frac{1}{x^2 + 1},$ which is entirely regular for real field values but contains singularities in the complex plane.}
			\label{fig:hill-pot}
\end{figure}
		
\begin{figure}[h]
			\centering
			\includegraphics[width=\linewidth]{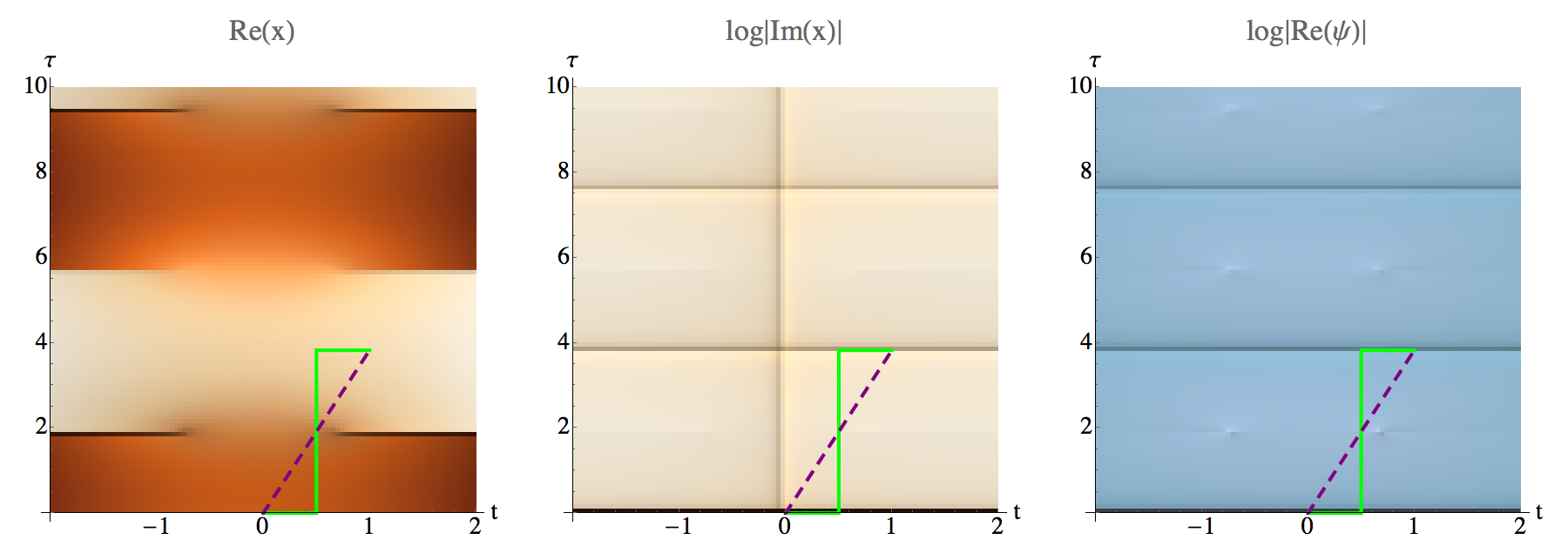}
			\caption{A first look at the solutions to the hill potential. The plots have been obtained with the initial condition that the particle is at rest at $x=-1.$ For a complete description, see the main text. The solid green and dashed purple lines show two equivalent tunneling paths, which will be further explored in Fig. \ref{fig:hill-pot-right-sing}.}
			\label{fig:hill-pot-right-sing-grid}
\end{figure}

\begin{figure}[h]
			\centering
			\includegraphics[width=\linewidth]{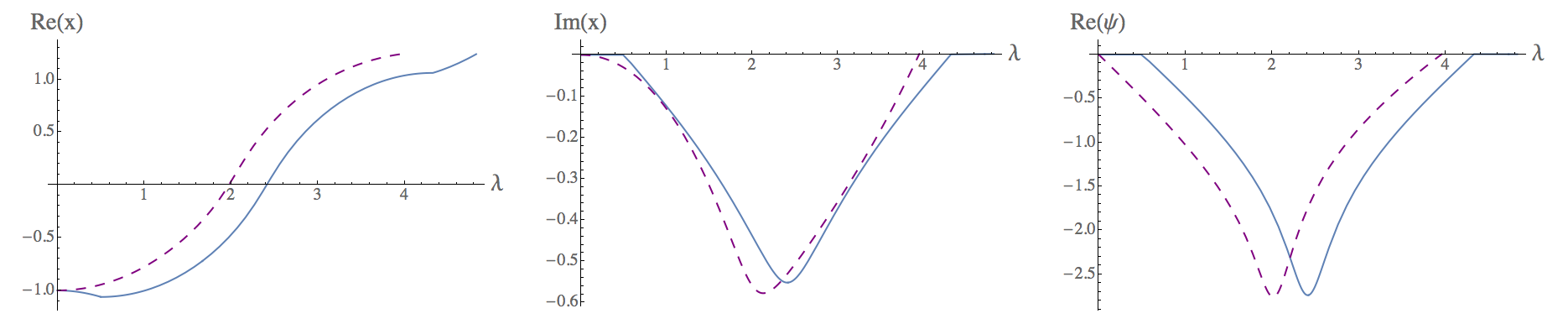}
			\caption{Solutions along the solid green (solid blue curve in this plot) and dashed purple (dashed purple line here) paths in Fig. \ref{fig:hill-pot-right-sing-grid}. These two paths are entirely equivalent to each other, since they link the same boundary values and since the region of the complex time plane that they enclose does not contain any singularity. The field evolutions along these paths are however different, as the graphs here show. Note also that the dashed purple path is neither Euclidean nor Lorentzian.}
			\label{fig:hill-pot-right-sing}
\end{figure}

For real $x$ values this potential is everywhere finite (see Fig. \ref{fig:hill-pot}), but in the complex plane there are singularities at $x(t)=\pm i.$ In classical physics these would not play any role, but in our treatment of quantum tunneling using complex time paths the singularities are important. They imply that there now exist possible tunneling paths that are distinct in the sense that they can encircle the singularities in various ways. It is then crucial to have a way of assessing which such paths truly contribute to the tunneling amplitude, and which do not. As in our previous discussion, we will approach this question by looking at the solutions of both the background and perturbation equations over extended regions in the complex time plane. For a first look see Fig. \ref{fig:hill-pot-right-sing-grid}. Given that we are now in the presence of singularities, we must be a little more precise in specifying how we obtained these figures. In Fig. \ref{fig:hill-pot-right-sing-grid} we have solved the equations of motion by taking paths that start at the original classical solution (on the real time axis), then run up the Euclidean time direction in between the two vertical lines of periodically spaced singularities (which can clearly be seen in the left panel, along with their attached outwards-running branch cuts), and from there branch out again parallel to the real time axis to the left and to the right. We see that in this way we can reach other classical solutions at periodically spaced lines of classicality parallel to the real time axis. Also, the right panel shows that nodes reside along the same lines. Thus we have a situation very similar to that of the simple inverted harmonic oscillator of section \ref{Sec:IHO}: adjacent lines of classicality may be joined by node-less, and thus relevant tunneling solutions, while tunneling paths between further separated lines necessarily cross at least one node and must be discarded. Thus the relevant paths pass just beyond the closest singularity right of the center. Two such paths (which are equivalent to each other) are shown in Figs. \ref{fig:hill-pot-right-sing-grid} and \ref{fig:hill-pot-right-sing} for illustration.
				
\begin{figure}[h]
			\centering
			\includegraphics[width=\linewidth]{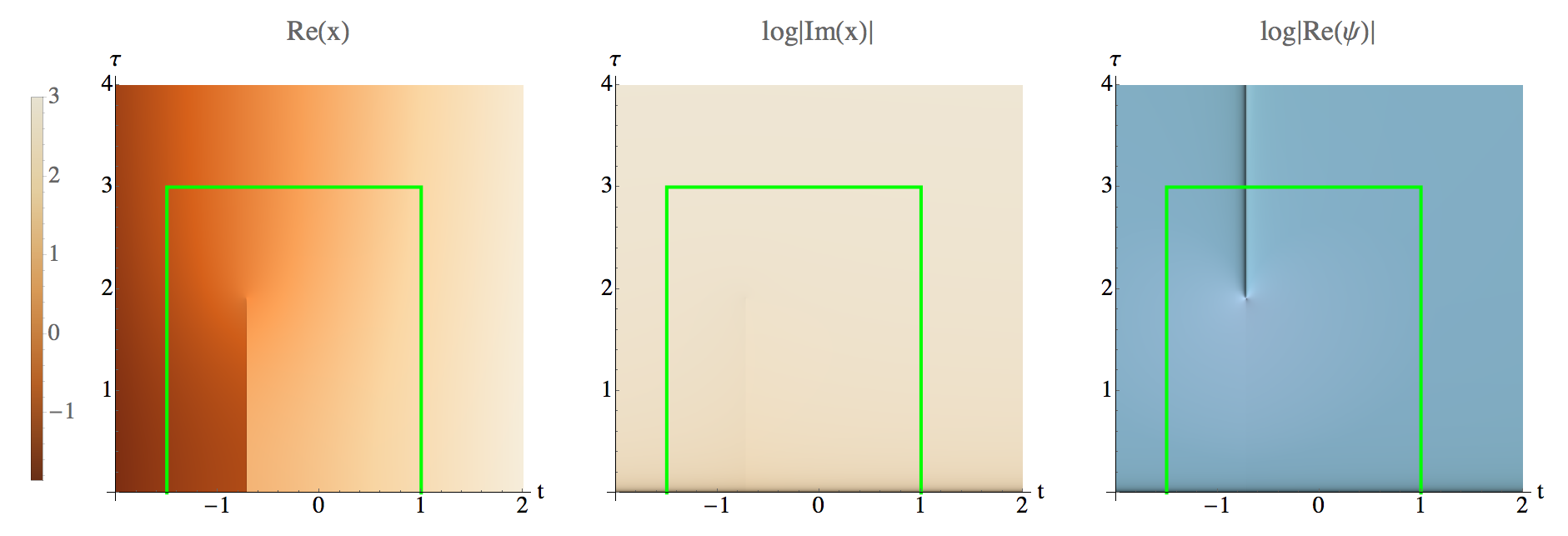}
			\caption{Investigating the left singularity. -- for a description of how this plot was obtained, see the main text. A new feature is the straight line of nodes emanating from the left singularity and running vertically upwards along the Euclidean time direction.}
			\label{fig:hill-pot-EoM-path-left-sing-grid}
\end{figure}

But now we have other possibilities too. In particular, we would like to know what happens when one chooses a path that passes by a singularity on the left. For this case, see Fig. \ref{fig:hill-pot-EoM-path-left-sing-grid}. Here we are choosing paths in the following manner: from the classical solution on the real time axis let the path run up on the left hand side of the closest singularity left of the center. Having passed that singularity, we continue parallel to the real time axis, and then branch out from there up and down along the Euclidean time direction. In practice this means that we have chosen the branch cut emanating from the singularity to run straight down perpendicular to the real time axis. We see something interesting: to the right of the branch cut, the solution becomes real again on the real time axis. Note that this real solution is now not reachable via purely classical evolution from the original classical solution on the left, because of the branch cut residing in between. However, the path circling around the left singularity is a possible tunneling solution linking these two classical solutions. Is it also a relevant one? The right panel shows that unfortunately this is not the case. There is a line of nodes starting at the singularity and running straight up -- any path joining the classical solution on the left to that on the right must necessarily intersect this line of nodes, and thus these solutions must all be discarded. One may wonder why the line of nodes is vertical in the present case. This is because the tunneling effectively occurs parallel to the real time axis, as opposed to the more usual situation where the tunneling is always along the Euclidean time direction. Here this occurs because of the presence of the singularity. As a consequence, near a node of a putative Euclidean solution the perturbation function could now be expanded in terms of $\sin(k t)$ functions, so that a line of nodes then emanates in the Euclidean time direction -- this is simply the rotated version of the argument presented around Eq. \eqref{node-location}.
	
\begin{figure}[h]
			\centering
			\includegraphics[width=\linewidth]{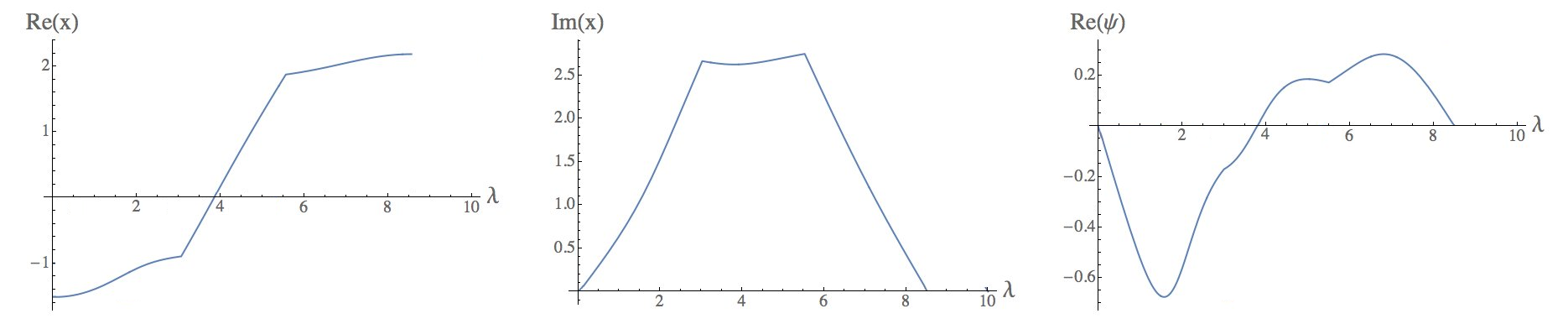}
			\caption{Solution along the green path in Fig. \ref{fig:hill-pot-EoM-path-left-sing-grid}.}
			\label{fig:hill-pot-EoM-path-left-sing}
\end{figure}
			
Other paths passing by singularities left of the center and further removed from the original classical solution will contain additional nodes, and hence all such paths are irrelevant. For the present potential, when we circle around the closest singularity left of the center by one additional full circle we essentially arrive back at the starting position, and thus no further paths need to be investigated. For other potentials, involving higher order singularities, additional non-trivial paths may exist, and our method will then allow one to determine all of the solutions relevant to quantum tunneling.

\section{Discussion}

Working in the semi-classical approximation, we have shown how complex time paths can mediate quantum tunneling between distinct classical histories. Both in order to find the location of the possible classical solutions and to determine the relevance of the solutions, we have shown that it is useful to solve the background and perturbation equations of motion over an extended region of the complexified time plane. This in particular enables one to find the nodes of the (real part of the) perturbation function, which, as we have argued, determine whether or not a given path contributes significantly to the tunneling amplitude. Our work extends previous treatments where single complex solutions have been considered. Moreover, our analysis of the perturbation function and its nodes is new. The latter analysis provides a crucial new aspect, with the absence of nodes being the criterion selecting the relevant paths.

It might be useful to add further comments contrasting our work with earlier approaches. The closest related works are those of Cherman-\"{U}nsal \cite{Cherman:2014sba} and Turok \cite{Turok:2013dfa}, which both aim to develop a description of quantum tunneling in ``real'' time, essentially by choosing a path in the complex time plane that is aligned as closely as possible with the real time axis. However, as our approach makes clear, although the contour can be chosen to be essentially aligned with the real time axis in some parts, the overall shift in Euclidean time is essential to capture tunneling. A special case is provided by the presence of singularities, in which case there may exist paths that encircle a singularity and then return back to the real line (though actually on a new sheet of the solution function), as exemplified in section \ref{Sec:Sing} -- still, at some point a departure from the real line is unavoidable to capture quantum effects.

Bender \cite{Bender:2008fr,Bender:2009zza} and Turok advocate using solutions with a complex energy to describe tunneling. In describing an initial wavepacket, this is in fact required, as emphasized by Turok. However, when describing a quantum transition between histories that can to a good approximation be described classically, there is no need to use complex energy solutions. In all our examples, we have chosen the energy to be real, as determined by the starting classical history. This is in no way an obstacle to describing tunneling by complex time paths. It is simply the initial conditions that determine the value of the energy. Note furthermore that since energy is conserved, a complex energy does not allow one to obtain a purely classical history after tunneling -- the best one can achieve is approximate classicality.

As shown by Cherman-\"{U}nsal and Turok, the imaginary part of the field may reach very large values during tunneling.  Turok has even proposed that these imaginary values may have a physical significance, and that they may be observable via weak measurements. We are skeptical about this claim, since the tunneling path may be deformed at will as long as one does not cross any singularities. Such deformations are allowed by Cauchy's theorem, and cannot result in any change in the physics. However, since the deformed paths reach different imaginary values of the field, these imaginary values cannot have a physical significance. It would however be fascinating if we were proven wrong about this point!

The advantage of our method is that it provides a rather general prescription for treating classical-to-quantum-to-classical transitions. This might be of great use in more complicated situations: we intend to extend our methods to quantum field theory, and also to semi-classical quantum gravity. In this case, one may generally expect singularities to be present and classical histories to come to an end, most notably near black hole or big bang type singularities (see \cite{Gielen:2015uaa} for work in this direction, and \cite{upcoming} for upcoming work). It is our (ambitious) hope that in such situations our method may be of use in identifying possible quantum transitions to other classical solutions.

\acknowledgements
We are thankful to Aleksey Cherman, Gerald Dunne and Mithat {\" U}nsal for useful critical comments.
We express our gratitude to the Max Planck Society for its support of the Theoretical Cosmology group at the Albert Einstein Institute.
G.L. acknowledges support from the Shota Rustaveli NSF Grant No.~FR/143/6-350/14.

\bibliographystyle{utphys}
\bibliography{Tunnelling}

\end{document}